\documentclass[a4paper,11pt]{article}
\pdfoutput=1 

\usepackage{jcappub} 

\usepackage[T1]{fontenc} 

\usepackage{graphicx}
\usepackage{color}

\def \be  {\begin{equation}}
\def \ee  {\end{equation}}
\def \ee  {\end{equation}}
\def \bea {\begin{eqnarray}}
\def \eea {\end{eqnarray}}

\def \l   {\lambda}

\newcommand{\nn}{\nonumber}

\renewcommand{\figurename}{{\bf Fig.}}
\renewcommand{\tablename}{{\bf Tab.}}

\def\be {\begin{equation}}
\def\ee {\end{equation}}
\def\bea {\begin{eqnarray}}
\def\eea {\end{eqnarray}}
\def\bc {\begin{center}}
\def\ec {\end{center}}
\def\bfg {\begin{figure}}
\def\efg {\end{figure}}
\def\bi {\begin{itemize}}
\def\ei {\end{itemize}}
\def\nn {\nonumber}

\def\la {\label}
\def\le {\left}
\def\ri {\right}

\def\fr {\frac}

\def\a  {\alpha}
\def\b  {\beta}

\def\D  {\Delta}

\def\l  {\lambda}

\def\m  {\mu}

\def\O  {\Omega}
\def\p  {\pi}

\def\t  {\tau}

\def\beq{\begin{equation}}
\def\eeq{\end{equation}}
\def\br{\begin{eqnarray}}
\def\er{\end{eqnarray}}
\newcommand{\eel}[1] {\label{#1}\end{equation}}

\title{\boldmath Impacts of Generalized Uncertainty Principle on Black Hole Thermodynamics and Salecker-Wigner Inequalities}

\author[a,b]{A.~Tawfik}
\affiliation[a]{Egyptian Center for Theoretical Physics (ECTP), MTI University, 11571 Cairo, Egypt}
\affiliation[b]{World Laboratory for Cosmology And Particle Physics (WLCAPP), Cairo, Egypt}
\emailAdd{a.tawfik@eng.mti.edu.eg}
\emailAdd{atawfik@cern.ch}

\abstract{We investigate the impacts of Generalized Uncertainty Principle (GUP) proposed by some approaches to quantum gravity such as String Theory and Doubly Special Relativity on black hole thermodynamics and Salecker-Wigner inequalities. Utilizing Heisenberg uncertainty principle, the Hawking temperature, Bekenstein entropy, specific heat, emission rate and decay time are calculated. As the evaporation entirely eats up the black hole mass, the specific heat vanishes and the temperature approaches infinity with an infinite radiation rate.
It is found that the GUP approach prevents the black hole from the entire evaporation. It implies the existence of remnants at which the specific heat vanishes. The same role is played by the Heisenberg uncertainty principle in constructing the hydrogen atom. We discuss how the linear GUP approach solves the entire-evaporation-problem.
Furthermore, the black hole lifetime can be estimated using another approach; the Salecker-Wigner inequalities. Assuming that the quantum position uncertainty is limited to the minimum wavelength of measuring signal, Wigner second inequality can be obtained.  If the spread of quantum clock is limited to some minimum value, then the modified black hole lifetime can be deduced. Based on linear GUP approach, the resulting lifetime difference depends on black hole relative mass and the difference between black hole mass with and without GUP is not negligible. }

\begin{document}
\maketitle
\flushbottom

\section{Introduction}

The idea of utilizing the fundamental limits governing mass and size of any physical system to register time dates back to nearly six decades. Salecker and Wigner were pioneers in suggesting the use a quantum clock \cite{wigner57,wigner58} in measuring {\it distances} between events in space-time. The measuring rods are entirely avoided, as they are supposed to be macroscopic objects \cite{wigner58}. This quantum clock is given as constrains on smallest accuracy and maximum running time as a function of mass and position uncertainties. In light of this, the Wigner second constrain is more severe than the Heisenberg uncertainty principle, which requires that only one single simultaneous measurement of both energy and time can be accurate. Wigner constrains assume that the repeated measurements should not disturb the clock. On other hand, the clock itself should be able to accurately register time over its total running period. 

The detectability of quantum space-time foam with gravitational wave interferometers has been addressed in Ref. \cite{disct1}, in which the authors criticized the limited measurability of the smallest quantum distances but gave an operative definition for the quantum distances and the elimination of the contributions from the total quantum uncertainty \cite{disct2}.  
Four decades later, Barrow applied Wigner inequalities to describe the quantum constrains on black hole lifetime \cite{barrow96}. It is found that the black hole running time should be correspondent to the Hawking lifetime, which is calculated under the assumption that the black hole is a black body and therefore the utilization of Stefan-Boltzmann law is eligible. Also, it is found that the Schwarzschild radius of black hole is correspondent to the constrains on Wigner size. Furthermore, the information processing power of a black hole is estimated by the emitted Hawking radiation \cite{swLitr}. 

The existence of a minimal length represents an exciting prediction from different approaches related to the quantum gravity such as String Theory. The mean idea is that the string is conjectured not to interact at distances smaller than its size, which is determined by its tension. For completeness, we add that information about the string interactions would be contained in the Polyakov action \cite{daxson}. This leads to generalizing Heisenberg uncertainty principle \cite{guppapers}. At Planck energy scale, the corresponding Schwarzschild radius becomes comparable to the Compton wavelength. Higher energies seem to result in further increase in Schwarzschild radius. In light of this, $\Delta x \approx \ell_{Pl}^2\Delta p/\hbar$. This observation and the ones deduced from the {\it gedanken} experiments would suggest that the GUP approach would essential at some concrete scales. 

The present work is organized as follows. The Generalized Uncertainty Principle is introduced in section \ref{sec:gup}. The connection of Heisenberg Uncertainty Principle with Hawking Thermodynamics is elaborated in section \ref{sec:1}. This sets an explanation for the catastrophic evaporation, section \ref{ce}. The consequences of the Generalized Uncertainty Principle on the black hole thermodynamics are studied in section \ref{sec:3}. A considerable consequence on the black hole Lifetime is analysed in section \ref{sec:bhl1}. Section \ref{sec:2} is devoted to the connection between 
Salecker-Wigner inequalities and the black hole lifetime. An additional estimation modified black hole lifetime due to GUP is outlined in section \ref{sec:SW_BHt}. The discussion and conclusions are presented in section \ref{sec:concl}.

\section{Generalized Uncertainty Principle}
\label{sec:gup}

The quadratic GUP approach was introduced in Ref.  \cite{guppapers,BHGUP1,BHGUP2,BHGUP3,BHGUP4,BHGUP5,BHGUP6,kmm,kempf,brau,Hossenfelder:2012jw}
\bea  \la{uncert1}
\Delta x_i \Delta p_i & \geq & \fr{\hbar}{2} \le[1 + \beta \le((\Delta p)^2 + <p>^2 \ri)  + 2\beta \le(\Delta p_i^2 + <p_i>^2\ri) \ri], 
\eea
where $p^2 = \sum\limits_{j}p_{j}p_{j}$ and $\beta=\beta_0/(M_{p}\, c)^2=\b_0 \ell_{p}^2/\hbar^2$. $M_{p}$ and $M_{p}\, c^2$ are Planck mass and energy, respectively.
It was shown that the inequality given in Eq. (\ref{uncert1}) is equivalent to the modified Heisenberg algebra \cite{kmm}
\be [x_i,p_j] = i \hbar ( \delta_{ij} + \beta \delta_{ij} p^2 +
2\beta p_i p_j ), \la{com1} \ee
which in turn assures, via the Jacobi identity, that
$[x_i,x_j]=0=[p_i,p_j]$ \cite{kempf}.\\

The presented approach for GUP seems to predict minimal measurable length and maximal observable momenta \cite{advplb,Ali:2010yn,Das:2010zf}. This is likely consistent with Doubly Special Relativity (DSR) theories, String Theory and Black Holes Physics 
\bea [x_i, p_j]\hspace{-1ex} &=&\hspace{-1ex} i
\hbar\hspace{-0.5ex} \left[  \delta_{ij}\hspace{-0.5ex} -
\hspace{-0.5ex} \alpha \hspace{-0.5ex}  \le( p \delta_{ij} +
\frac{p_i p_j}{p} \ri)
+ \alpha^2 \hspace{-0.5ex}
\le( p^2 \delta_{ij}  + 3 p_{i} p_{j} \ri) \hspace{-0.5ex} \ri], 
\label{comm01}
\eea
where
$\alpha = {\alpha_0}/{M_{p}c} = {\alpha_0 \ell_{p}}/{\hbar}$ and  $\ell_{p}$ is Planck length.
The DSR transformations preserve both speed of light and  invariant energy scale \cite{sm}. It is not surprising that Eq. (\ref{comm01}) imply the existence of minimum measurable length and maximum measurable momentum
\bea
\D x &\geq& (\D x)_{min}  \approx \alpha_0\ell_{p}, \la{dxmin}\\
\D p &\leq& (\D p)_{max} \approx \frac{M_{p}c}{\alpha_0}. \la{dpmax}
\eea
Furthermore, the proposed GUP suggests that the space is discrete \cite{advplb,Ali:2010yn,Das:2010zf} and that all measurable lengths are quantized in units of a fundamental minimum measurable length (which can be the Planck length). It is notable that a similar quantization of the length was shown in context of Loop Quantum Gravity \cite{LQG}.

Since the GUP approach modifies the fundamental commutator bracket between position and momentum, then the Hamiltonian runs to be modified, nationally. Therefore, the GUP approach should possess effects on different quantum phenomenon. In light of this, it is obviously important to study  such effects. Their quantitative estimations would open a window to the quantum gravity phenomenology. In a series of papers, the author collaborating or individually has investigated the effects of the linear GUP approach on compact stars \cite{Ali:2013ii}, Newtonian law of gravity \cite{Ali:2013ma}, inflationary parameters and thermodynamics of the early universe \cite{Tawfik:2012he}, Lorentz invariance violation \cite{Tawfik:2012hz} and measurable maximum energy and minimum time interval \cite{DahabTaw}. 
Furthermore, the effects of quantum gravity on the quark-gluon plasma are studied  \cite{Elmashad:2012mq}. It was found that the GUP can potentially explain the small observed violations of the weak equivalence principle in neutron interferometry experiments \cite{exp} and also predicts a modified invariant phase space which is relevant to LT. It is suggested \cite{Pikovski:2011zk} that GUP can be measured directly in Quantum Optics Lab which apparently confirms the theoretical predictions given in Ref. \cite{dvprl,Ali:2011fa}.

The present paper is organized is three parts. The first part introduces Hawking thermodynamics in connection with the Heisenberg uncertainty principle. We briefly introduce how the thermodynamic quantities, Hawking temperature, Bekenstein entropy, specific heat, emission rate, and decay time can be obtained from the standard uncertainty principle in $D$- dimensions. In Sec \ref{sec:3}, we investigate the impact of GUP, Eq.\ (\ref{comm01}), on black hole thermodynamics. The black hole lifetime can be estimated from the Salecker-Wigner Inequalities, which will be outlined in the third part, section \ref{sec:2}.

\section{Uncertainty Principle and Hawking Thermodynamics}
\label{sec:1}

The connection between standard Hawking temperature and Heisenberg uncertainty principle has been introduced  by Adler {\it et al.} \cite{Adler}, where the black hole is assumed to behave as a black body radiator. Cavaglia {\it et al.} \cite{Cavaglia:2003qk,Cavaglia1,Cavaglia2} generalized the relation in large extra dimensions. A black hole can be modelled as $(D-1)$-dimensional sphere of size equal to twice of  Schwarzschild radius, $r_s=2 G m/c^2$. For the emitted particles,
\be 
\Delta p_i \Delta x_j \geq \frac{\hbar}{2}
\delta_{ij}. \label{SUP} 
\ee
Consequently, an emitted Hawking particle has a minimum value of position uncertainty 
\be \Delta x \approx 2 r_s =2 \lambda_D
\le[\frac{G_DM}{c^2}\ri]^{1/(D-3)}\label{DU}, \ee
where 
\bea
\l_D &=& \le[\frac{16\p}{(D-2)\O_{D-2}}\ri]^{1/(D-3)}, \\
\O_D &=& \frac{2 \pi^{\frac{D-1}{2}}}{\Gamma(\frac{D-1}{2})}, 
\eea
and $G_D$ is the gravitational constant in $D$-dimensional space-time. From Eqs. (\ref{SUP}) and (\ref{DU}) and the argument used in Ref. \cite{Cavaglia:2003qk,Cavaglia1,Cavaglia2} that $\Delta x_i \Delta p_i \approx \Delta x \Delta p$ , then the emitted  Hawking particle is assumed to have an energy uncertainty 
\be 
\D E \approx 
\frac{M_{p}c^2}{4\l_D}\le(\frac{M}{M_{p}}\ri)^{-1/(D-3)}.
\label{EU} 
\ee
From now on, we set $m=M/M_p$ i.e., the mass of black hole is normalized to the Planck mass $M_p=(\hbar^{D-3}/c^{D-5}G_D)^{1/(D-2)}$. 
The energy uncertainty $\D E$ can be identified as the energy of the emitted
photon \cite{Adler}. Based on this, the characteristic temperature of  Hawking particle in D-dimensions of space-time \cite{Hawking:1974sw} can be obtained from $\D E$ simply through multiplying it with a calibration factor $(D-3)/\pi$ \cite{Adler,Cavaglia:2003qk,Cavaglia1,Cavaglia2} 
\be
T_H=\frac{D-3}{4\p\l_D}~M_{p}c^2~m^{-1/(D-3)}\,.
\la{hawT}
\ee
The entropy is to be calculated from the first law of thermodynamics
\be
dM=\frac{T}{c^2}\, d S.
\label{TD}
\ee
In terms of $m$,  Eq.\ (\ref{TD}) can be rewritten as 
\be 
d S= \frac{M_p\, c^2 }{T}\, dm. \la{TD1} 
\ee 
Integrating Eq.\ (\ref{TD1}) and using Eq. (\ref{hawT}) leads to Bekenstein
entropy \cite{Bekenstein:1973ur} 
\be
S=\frac{4 \pi\, \l_D}{D-2}\, m^{(D-2)/(D-3)}.\la{entropy}
\ee
From Eqs.\ (\ref{TD1}) and (\ref{hawT}), the specific heat reads
\bea 
{\cal C} &=& T \frac{\partial S}{\partial T}
= -4 \pi\, \l_D\, m^{\frac{(D-2)}{(D-3)}}, \la{C0} 
\eea

The Hawking temperature $T_H$ can be used to estimate the emission rate. If a black body radiator is assumed and the energy loss is dominated by photons, then Stefan-Boltzmann law gives a straightforward estimation for the emission rate. For instance, when assuming a $D$-dimensional space-time brane, the thermal emission in the bulk of brane with mass $m$ can be neglected and the black hole is supposed to radiate mainly on the brane \cite{evaporation} so that the emission rate can be
given as
\be
\frac{d M}{d t} \propto T^{D}.
\label{dMdt}
\ee
Based on these assumptions, the brane would exist in $D=4$ and its mass decay reads
\be
\frac{d m}{d t}=-\frac{\m^{\prime}}{t_{p}}~m^{\frac{-2}{(D-3)}}\,,
\la{rate1}
\ee
where $t_{p}=(\hbar\, G_D/c^{D+1})^{1/(D-2)}$ is the Planck time. An expression for $\mu^{\prime}$ can be found in Ref. \cite{Cavaglia:2003qk,Cavaglia1,Cavaglia2}.
Integrating Eq.\ (\ref{rate1}) results in the decay time of the black hole
\be 
\t =  \m^{\prime -1} \le(\frac{D-3}{D-1}\ri) m_i^{\frac{(D-1)}{(D-3)}}~t_{p}, \la{decayt} 
\ee 
which is directly related to Planck time and black hole mass.

It is apparent that in all quantities, Hawking temperature $T_H$, Bekenstein entropy $S$, specific heat ${\cal C}$, emission rate $dm/dt$, and decay time $\t$ lead to catastrophic evaporation as the mass $m$ radiates. The specific heat vanishes only when the mass vanishes. Consequently, the black hole is supposed to continue radiating until $m=0$. But as the mass approaches zero, its temperature approaches infinity with infinite radiation rate.

\subsection{Catastrophic Evaporation}
\label{ce}

It is believed that the {\it ''catastrophic evaporation''} would be an artifact stemming from extending the approximation that Hawking made. Hawking approximation neglects the mass-loss of the black hole during evaporation, beyond its limits of applicability. If the mass of the black hole becomes comparable to its temperature, which is the case when the mass reaches the Planckian regime, then it is no longer appropriate to use the macro-canonical ensemble. Instead, the micro-canonical ensemble has to be used.  Recently, this has been nicely summarized \cite{revw}. It claims to remove weaknesses of the standard thermodynamic description of black holes, macro-canonical. In other words, the intensivity and extensivity of measured quantities are thought to reflect the statistical aspects related to the use of canonical and grand-canonical ensemble \cite{TawMBXXXI}.

Furthermore, from the quantum back reaction due to $N$ massless fields that may be worked out to a considerable detail in a variant of integrable dilation gravity model in two dimensions, a critical mass of collapsing object of order $\hbar\, N\, \Lambda^{1/2}$ would remain \cite{cr1}, where $\Lambda$ is the cosmological constant. Above it, the end point of Hawking evaporation is two disconnected remnants of infinite extent, each separated by a mouth from the outside region. Deep inside the mouth there is a universal flux of radiation in all directions, in a form different from Hawking radiation. Below the critical mass, no remnant is left behind implying complete Hawking evaporation or even showing no sign of Hawking radiation.

\section{Generalized Uncertainty Principle and Black Hole Thermodynamics}
\label{sec:3}

As presented in Ref. \cite{afa:jhep1}, starting from Eq. (\ref{comm01}), in which the arguments introduced in Ref.  \cite{Cavaglia:2003qk,Cavaglia1,Cavaglia2} are implemented, a general $(D-1)$-dimensional inequality  can be derived as follows. 
 \bea \la{ineq}
 \D x \D p &\geq&
 \frac{\hbar}{2}\le[1- \a \le\langle p \ri\rangle- \a \le\langle
 \frac{p_i^2}{p} \ri\rangle
+\a^2 \le\langle p^2 \ri\rangle + 3 \a^2 \le\langle p_i^2\ri\rangle \ri], \nonumber \\  
  &\geq& \frac{\hbar}{2} \le[1- \a_0 ~\ell_p~
\le(\frac{4}{3}\ri)~\sqrt{\mu}~~ \frac{\D p}{\hbar}+ ~2~
(1+\mu)~ \a_0^2 ~\ell_p^2 ~ \frac{\D p^2}{\hbar^2} \ri].
\la{ineqII} 
\eea 
To our knowledge, this inequality is the only one that directly follows from Eq.\ (\ref{comm01}).
It can be solved in $\D p$, 
\be 
\frac{\D p}{\hbar}\geq\frac{2 \D
x+\a_0
~\ell_p~\le(\frac{4}{3}\ri)~\sqrt{\mu}}{4~(1+\mu)~\a_0^2~\ell_{p}^2}\le[1-
\sqrt{1-\frac{8~(1+\mu)~\a_0^2\ell_{p}^2} {\le(2 \D x+\a_0
\ell_p\le(\frac{4}{3}\ri) ~\sqrt{\mu}~\ri)^2}}~\ri],\la{gupso} 
\ee
where $\mu=[2.821 (D-3)/\pi]^2$. 
Only the solution with the negative sign is considered to be physical and assumed to fulfil the standard uncertainty relation at $\ell_p/\D x \rightarrow 0$.

As given in Sec. \ref{sec:1}, a modified Hawking temperature can be deduced
\bea  \la{modT}
T_H^{\prime}
&=& \frac{D-3}{\pi\, \a_0^2} \frac{M_p\, c^2}{(1+\mu)}\le(\l_D\, m^{1/(D-3)} + \frac{\a_0~
\sqrt{\mu}}{3}\ri)
\le[1-\sqrt{1-\frac{(1+\mu)~\a_0^2 }{2 \le(\l_D m^{1/(D-3)}+\frac{\a_0\sqrt{\mu}}{3}\ri)^2}}\ri], \nn \\
&=& 2 T_H \le(1+\frac{\a_0 ~\sqrt{\mu}~}{3~\l_D
m^{\frac{1}{D-3}}}\ri)^{-1} \le[1+\sqrt{1-\frac{(1+\mu)~ \a_0^2
}{2 \le(\l_D m^{1/(D-3)}+ \frac{\a_0 ~\sqrt{\mu}}{3}
\ri)^2}}\ri]^{-1}, \eea
which is considered as physical as long as the mass of the normalized black hole mass and $\alpha_0$ satisfy the inequality
\be 
(1+\mu)~\a_0^2  \leq  2 \le(\l_D
m^{1/(D-3)}+\frac{\a_0~ \sqrt{\mu}}{3}~\ri)^2.
\ee 
This inequality tells us that the black hole should keep a minimum mass $M_{min}$ not evaporated
\be
M_{min}=M_p \le(\sqrt{\frac{(1+\mu)}{2}}-\sqrt{\frac{\mu}{9}}\ri)^{D-3} \frac{D-2}{8 \Gamma\le({\frac{D-1}{2}}\ri)}
\le(\a_0 \sqrt{\pi}\ri)^{D-3}. \la{MinMI}
\ee
It is worthwhile to notice here that the minimum mass of the black hole looks different from the
one obtained in Ref. \cite{Cavaglia:2003qk,Cavaglia1,Cavaglia2}. On other hand, the linear GUP approach \cite{advplb,Ali:2010yn,Das:2010zf} generates the factor  $\le(\sqrt{(1+\mu)/2}-\sqrt{\mu/9}\ri)^{D-3}$, which obviously increases the value of the minimum (remnant) black hole mass.

\subsection{Black Hole Lifetime}
\label{sec:bhl1}

If GUP approach is implemented, the Hawking evaporation seems to stop when the size of the black hole is given by the Planck length and the temperature approaches the maximum value 
\bea T_{max} \approx 2
\le[\frac{\frac{3(1+\mu)}{2}+\sqrt{\frac{\mu(\mu+1)}{2}}}{\frac{3}{2}
+ \frac{7}{6} \mu}\ri]~ T_H. \eea
As discussed previously, the emission rate can be calculated using Stefan-Boltzmann law. From  Eqs.  ({\ref{dMdt}) and (\ref{rate1}}). The decay rate for a 4-dimensional brane reads
\be
\frac{dm}{dt}= -16  \frac{\mu^{\prime}}{t_p} m^{\frac{-2}{D-3}} \le(1+\frac{\a_0~\sqrt{\mu}}{3\l_D m^{\frac{1}{D-3}}}\ri)^{-4}
\le[1+\sqrt{1-\frac{(1+\mu) \a_0^2 }{2 \le(\l_D m^{\frac{1}{D-3}}+\frac{\a_0\sqrt{\mu}}{3}\ri)^2}}\ri]^{-4}.
\ee

Furthermore, the entropy and specific heat, respectively,  are given as
\bea
dS &=& \frac{2\pi}{D-3}\, \l_D \, m^{\frac{1}{D-3}} \le(1+\frac{\a_0~\sqrt{\mu}}{3 \l_D \, m^{\frac{1}{D-3}}}\ri)
\le[1+\sqrt{1-\frac{(1+\mu)\, \a_0^2 }{2\, \le(\l_D\, m^{\frac{1}{D-3}}+\frac{\a_0\, \sqrt{\mu}}{3}\ri)^2}}\ri]~dm, \hspace*{10mm}\\
{\cal{C}}&=&-\frac{2 \pi}{\l_D}\, m^{\frac{D-4}{D-3}} \; \le(\l_D\, m^{\frac{1}{D-3}}+ \frac{\a_0~\sqrt{\mu}}{3} \ri)^2
\sqrt{1-\frac{(1+\mu)\, \a_0^2 }{2\, \le(\l_D\, m^{\frac{1}{D-3}}+\frac{\a_0\, \sqrt{\mu}}{3}\ri)^2}}\nn\\&&
\le[1+\sqrt{1-\frac{(1+\mu) \a_0^2 }{2\, \le(\l_D m^{\frac{1}{D-3}}+\frac{\a_0\, \sqrt{\mu}}{3}\ri)^2}}\ri].
\eea

We notice that the GUP-approach prevents black holes from entire evaporation.  The GUP implies the existence of black hole remnants at which the specific heat vanishes. In other words, it is not necessary that the mass entirely vanishes in order to assure that the specific heat vanishes. Reaching this stage, the black hole is no longer able to exchange heat with surrounding space.

The atomic physics gives a similar example. The Heisenberg uncertainty principle plays a similar role in constructing the hydrogen atom. It prevents it for collapse \cite{Adler,Cavaglia:2003qk,Cavaglia1,Cavaglia2}.

There is another approach that can be utilized in order to estimate the black hole lifetime. Salecker-Wigner inequalities, section \ref{sec:2}, would play a role in estimating the lifetime of black holes.
 
\section{Salecker-Wigner Inequalities and Black Hole Lifetime}
\label{sec:2}

As anticipated in the introduction, the second Wigner inequality is more severe than the standard Heisenberg energy-time uncertainty principle. This is simply because it requires that a quantum clock is able to show proper time even after the time was being read. In other words, the quantum uncertainty in its position does not produce a significant inaccuracy in its time measurement. This property is conjuncted to hold over long periods i.e., coherent time intervals. The terminology ''coherence" has to do with the correlation properties of the signal used in the measurement. The ''coherent time'' is defined as the time period within which the signal remains ''coherent''.
\bea
\tau_{c} &=& \frac{1}{\D\, \nu_c} \approx \frac{\lambda_c^2}{c\, \D \lambda_c},
\eea
where the subscript $c$ refers to coherence.

From Heisenberg uncertainty principle, the momentum uncertainty in {\it single analogue} quantum  clock of mass $m$  is $\hbar/2 \D x$, where $\D x$ is uncertainty in its quantum position. After time $t$, the clock position spread increases to 
\bea \label{eq:ts}
\D x^{\prime} &=& \D x + \frac{\hbar\, t}{m}\, \frac{1}{2 \D x}. 
\eea
Assuming that the mass of quantum clock remain unchanged, then Eq. (\ref{eq:ts}) leads to a minimum time spread   
\bea \label{eq:ts2}
\D x &\geq& \sqrt{\hbar\, \frac{t_{max}}{2\, m}},
\eea
where $t_{max}$ is the total ''coherent'' time.  Expression (\ref{eq:ts}) is known as Wigner first inequality. 

In the case that the mass depends on the uncertainty in position, then the minimum time spread  reads 
\bea
\D x &\geq& \frac{\hbar\; t_{max}\, m^{\prime} - \sqrt{\hbar\; t_{max}\, \left[8\, m^2 + (m^{\prime})^2\, \hbar\, t_{max}\right]}}{4\, m^2},
\eea
where $m^{\prime}=d m/d \D x$. As undertaken in Eq. (\ref{gupso}), the positive sign is evaluated as non-physical.
 
If the time measurements are repeated and they have to remain reliable, then the position  uncertainty which in turn must be caused by the repeated measurements, should be smaller than the minimum wavelength of the reading signals i.e., $\D x \leq c\, T_{min}$. For an unsqueezed, unentangled and Gaussian signal, the minimum size can be give in minimum mass of the quantum clock. From Eq. (\ref{eq:ts2}),  the mass-time inequality reads
\bea \label{eq:ts22}
m &\geq& \frac{\hbar}{2\, c^2}\; \frac{t_{max}}{t_{min}^2},
\eea
which is known as Wigner second inequality.

\subsection{Black Hole Lifetime}
\label{sec:SW_BHt}

Assuming a black hole of a size comparable to the Schwarzschild radius, $r_s=2\, G\, m /c^2$, then  Wigner first inequality, Eq. (\ref{eq:ts}), can be applied on it. From Eq. (\ref{eq:ts2}), the maximum running time (lifetime) of black hole reads
\bea
t_{max} & \leq & 8 \frac{G^2\, m^3}{\hbar \, c^4},  \label{eq:bhlt1}\\
 & \leq & 8\, \frac{G}{c^3}\left(\frac{m^3}{M_p^2}\right),  \label{eq:bhlt2}
\eea
where $M_p=\sqrt{c \hbar/G}$ is Planck mass. Here after, we refer to black hole mass as $m$. It should be mixed with the normalized mass of black hole mentioned in previous sections.  Obviously, these expressions are compatible with the Hawking lifetime \cite{hawlt}. Eqs. (\ref{eq:bhlt1}) and (\ref{eq:bhlt2}) give answers to the question, {\it ''how does the life of a black hole run out?''}. As discussed in the previous sections, the mass of black hole quantum clock is the only parameter that describes a reliable mechanism.  It offers an alternative possibility not based on the assumption that black hole has to be a black body radiator \cite{hawlt}, as anticipated sections \ref{sec:1} and \ref{sec:3}.

At Planckian scale, the space-time fluctuation become significant. Therefore, it is natural to set a bound to the linear spread of the quantum clock, Eq. (\ref{eq:ts2}). The natural bound is the Planck distance. As given in introduction, the GUP-approach gives prediction for a minimal measurable length, Eq. (\ref{dxmin}). Therefore, $\alpha_0\, \ell_{p}$ would be taken as the smallest linear spread of the quantum clock.

For a quantum clock having a position uncertainty $\D x$, its momentum uncertainty can directly be deduced from Eq. (\ref{gupso}). At time $t$, the position uncertainty due to GUP becomes
\bea
\D x^{\prime} &=& \D x + \frac{2 \D x + \frac{4}{3}\, \a_0
~\ell_p~\sqrt{\mu}}{4~(1+\mu)~\a_0^2~\ell_{p}^2\, m} \, \hbar \, t \, \le[1-
\sqrt{1-\frac{8~(1+\mu)~\a_0^2\ell_{p}^2} {\le(2 \D x + \frac{4}{3}\, \a_0
\ell_p~\sqrt{\mu}~\ri)^2}}~\ri].
\eea
Then 
\bea
\D x_{GUP} &\geq & \frac{1}{2} \left[-A_1 + \frac{\sqrt{2}\, (m\, A_2 + 2\, \hbar\, t)^2}{\sqrt{m\,(m\, A_2 + 2\, \hbar\, t)^2\; (m\, A_2 + 4\, \hbar\, t)}}\right], \label{eq:Dx1}
\eea
where
\bea
A_1 &=&  \frac{4}{3} \alpha_0 \ell_p \sqrt{\mu}, \\ 
A_2 &=& 4 (1+\mu)   \alpha_0^2 \ell_p^2. 
\eea
At $\alpha_0=0$ i.e., switching off the GUP effects, Salecker-Wegner position uncertainty is recovered
\bea
\D x_{SW} &\geq & \sqrt{\hbar \frac{t}{2\, m}}.  \label{eq:Dx2}
\eea
In Eqs. (\ref{eq:Dx1}) and  (\ref{eq:Dx2}) the negative solutions are evaluated as non-physical. It is apparent that Eq. (\ref{eq:Dx2}), in which GUP effects are excluded, is identical with the Wigner first inequality Eq. (\ref{eq:ts2}). The difference between Eq. (\ref{eq:Dx1}), in which GUP effects are taken into account, and Eq. (\ref{eq:ts2})  simply reads
\bea
\D x_{GUP}-\D x_{SW} &=& \frac{1}{2}\left[-A_1-\hbar^2\, t_{max}^2 \sqrt{\frac{2}{m\, \hbar^3\, t_{max}^3}} \right. \nonumber \\
&& \left. \hspace*{5mm}\; + (m\, A_2 + 2 \hbar\, t_{max})^2 \sqrt{\frac{2}{m(m\, A_2 + 2 \hbar t_{max})^2 (m\, A_2 + 4 \hbar t)}}\right], \hspace*{10mm}
\eea
which obviously vanishes at vanishing $\alpha_0$.

\begin{figure}[tbp]
\includegraphics[width=.95\textwidth]{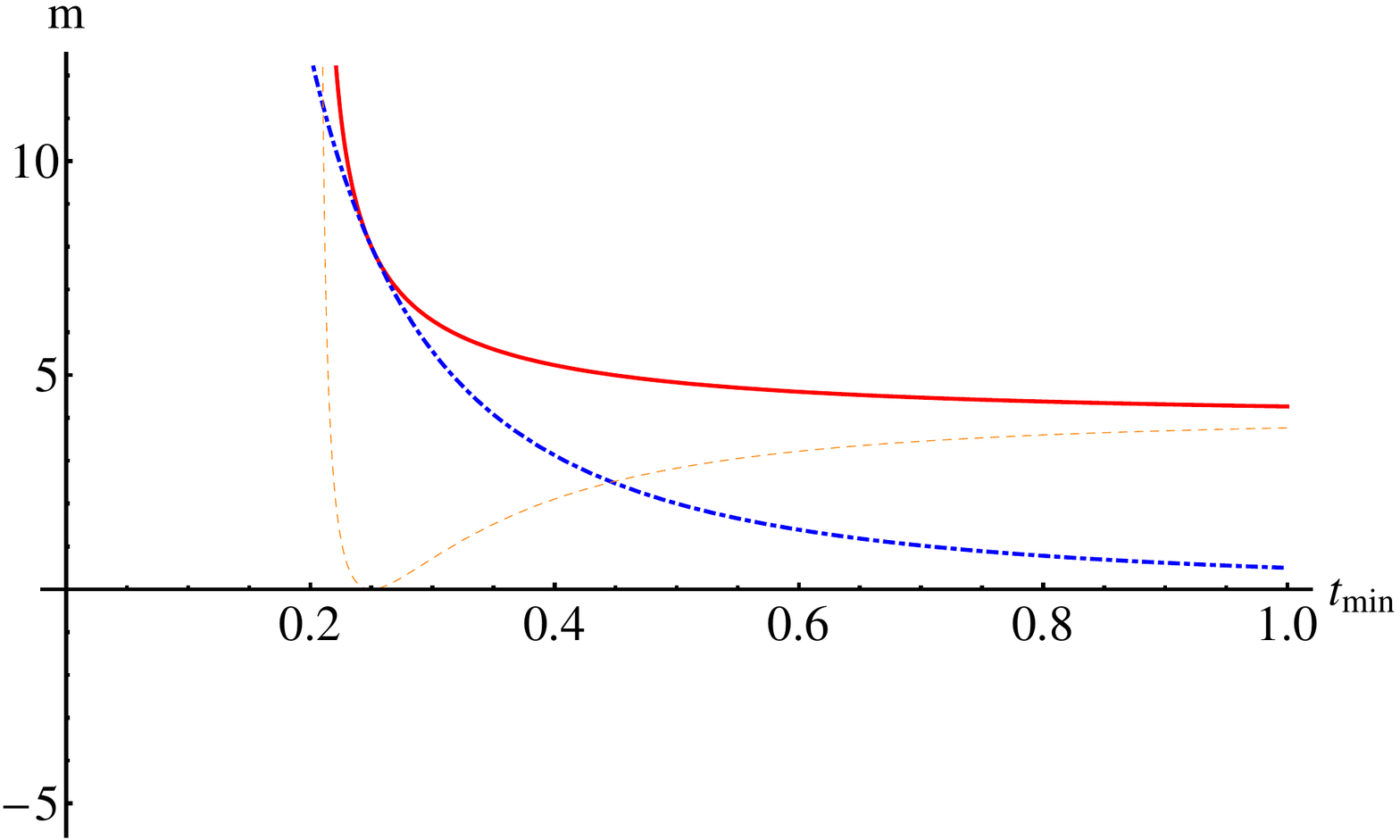}
\caption{The mass of black hole is given in dependence on its lifetime with (solid line) and without (dash-dotted line) GUP and their difference (dashed line).  The values of the variables $A_1$, $A_2$, $\hbar$, and $c$ are taken unity. }
\label{fig:tmax1}
\end{figure}

Assuming that the quantum position uncertainty should not be larger that the minimum wavelength of measuring signal, so that in Eq. (\ref{eq:Dx1}), we set $\D x_{GUP}\leq c\, t_{min}$,
\bea \label{eq:Wm22}
m_{GUP} &\geq & \frac{-\left[2\, \hbar \, t_{max} A_3 \pm 2 \hbar\, t_{max} \left(A_1+2 c\, t_{\min}\right) \sqrt{
A_3}\right]}{A_2\, A_3},
\eea  
where $A_3 = A_1^2-2 A_2+4 c\, t_{\min} \left(A_1+c \, t_{\min}\right)$. The positive sign defines a non-physical solution, where 
\bea
2 \hbar\, t_{max} \left(A_1+2 c\, t_{\min}\right) \sqrt{A_3} &>& 2\, \hbar \, t_{max} A_3,
\eea
implying that 
\bea \label{eq:a3}
\sqrt{A_3} & < & A_1+2 c\, t_{\min}.
\eea
At vanishing $\alpha_0$, Eq. (\ref{eq:Wm22}) goes back to Wigner second inequality, Eq. (\ref{eq:ts22}). At this scale, the inequality, Eq. (\ref{eq:a3}), turns in an equality in $t_{min}$. The difference between Eq. (\ref{eq:Wm22}) and Eq. (\ref{eq:ts22}) results in
\bea
m_{GUP}-m_{SW} &=& \frac{1}{2} \left(-\frac{4\, \hbar\, t_{max}}{A_2} - \frac{\hbar t_{max}}{c^2\, t^2_{min}}  -\frac{4\, \hbar^2\, t_{max}^2 (A_1 + 2\, c\, t_{min})^2}{A_2\,\sqrt{\hbar^2\, t_{max}^2\, (A_1 + 2\, c\, t_{min})^2\, A_3}} \right).
\eea

The modified black hole lifetime can be derived assuming that the spread of quantum clock has a minimum value, the Schwarzschild radius, $r_s$,
\bea \label{eq:tmbh}
t_{GUP} &=& \frac{1}{16\, \hbar^2}\left[-\hbar\, m\, A_4 - \hbar\, m\, A_4 \left(1-128\,  A_2\right)^{1/2}\right],
\eea
where $A_4=-4 A_1^2+8 A_2-16  r_s A_1-16 r_s^2$. The solution including negative sign is taken as physical. At $\alpha_0=0$, the modified black hole lifetime, Eq. (\ref{eq:tmbh}), goes back to Wingner inequality, Eq. (\ref{eq:bhlt1}). The difference between black hole lifetime in GUP-approach and Wingner inequality reads
\bea
t_{GUP}-t_{SW} &=& 2 \frac{m\, r_s^2}{\hbar}=8\frac{G\, m}{c^3}\, \left(\frac{m}{M_p}\right)^2,
\eea
and depicted in Fig. \ref{fig:tmax1}.

\begin{figure}[tbp]
\includegraphics[width=.95\textwidth]{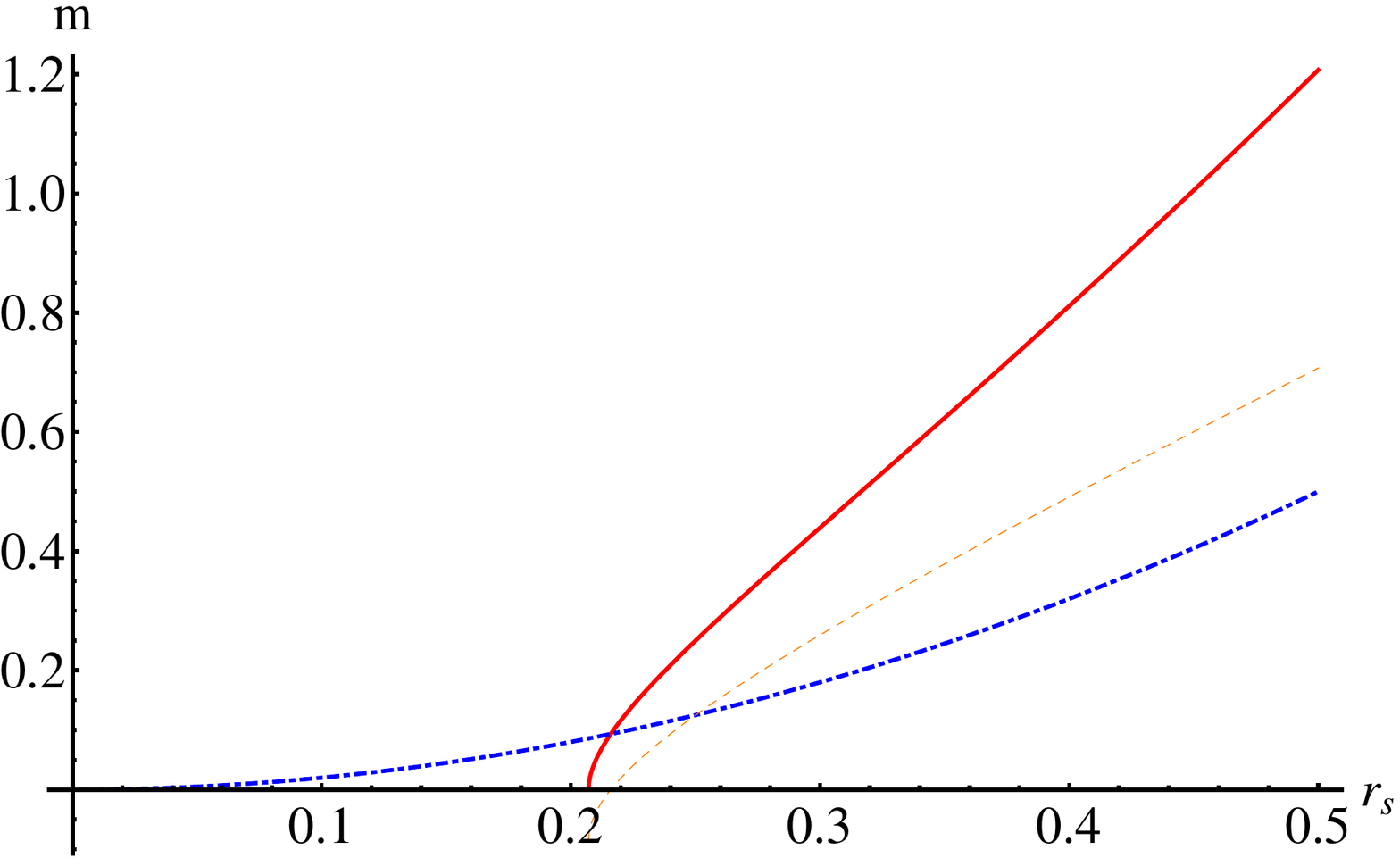}
\caption{The mass of black hole mass is given in dependence on $r_s$. The values of the variables $A_1$, $A_2$, $\hbar$, and $c$ are taken unity. }
\label{fig:tmax2}
\end{figure}

Employing GUP-modification of Salecker-Wigner inequalities in order to find the modified black-hole lifetime has been discussed in literature \cite{swLitr}. The authors of Ref. \cite{swLitr} utilized a quadratic GUP approach, especially the one that was suggested by Scardigli \cite{BHGUP4}.  There are essential differences between the difference GUP approaches. Table \ref{table1} summarizes a short comparison between quadratic \cite{guppapers,BHGUP1,BHGUP2,BHGUP3,BHGUP4,BHGUP5,BHGUP6,kmm,kempf} and linear \cite{Ali:2010yn,Das:2010zf} GUP approaches in different aspects, namely Heisenberg algebra, minimal length uncertainty and maximal moment uncertainty and maximal measured moment. An extensive comparison is given in Ref. \cite{TD2013}. In the present work, we utilize a linear approach \cite{Ali:2010yn,Das:2010zf}. Accordingly, the mass dependence of modified black hole lifetime  obtained in Ref.  \cite{swLitr} obviously differs from the one reported in the present work, Eq. (\ref{eq:tmbh}). Such a difference reflects the essential improvements in constructing the linear GUP approach \cite{Ali:2010yn,Das:2010zf}. 

Furthermore, correction terms for temperature, entropy, and heat capacity of Schwarzschild black hole are obtained from the commutation relations in the framework of modified uncertainty principle suggested by DSR \cite{Zeynali}. The authors utilized quadratic GUP  \cite{kmm,kempf} and GUP-DSR \cite{gupdsr} approaches. It is found the latter considerably delays the black hole evaporation. Therefore, mass, entropy and specific heat last longer relative to the quadratic GUP. With this regard, an extensive comparison between linear GUP and GUP-DSR should be undertaken. The GUP-DSR implements Heisenberg algebra, in which the commutation relation is given in dependence on a linear combination of Planck length of first and second order. 

\begin{table}[htb]
\begin{center}
\begin{tabular}{| l || l | l |}
\hline \hline
  & Quadratic GUP \cite{guppapers,BHGUP1,BHGUP2,BHGUP3,BHGUP4,BHGUP5,BHGUP6,kmm,kempf} & Linear GUP  \cite{Ali:2010yn,Das:2010zf} \\ 
\hline  \hline
Heisenberg Algebra & $\left[x,p\right]= i \hbar \left(1+\beta p^{2}\right)$ & $\left[x,p\right]= i \hbar \left(1-\alpha p +2 \alpha ^{2} p^{2} \right)$\\ 
\hline 
Minimal length uncertainty $\Delta x$ & $ \hbar \sqrt{\beta} $ & $ \hbar \alpha $    \\ 
\hline 
Maximal moment uncertainty $\Delta p$ & Undetermined & $M_{pl} c/\alpha _{0} $ \\ 
\hline 
Maximal moment $P_{max} $ & Divergence  & $1/(4\alpha)$   \\  
\hline 
& & String Theory  \\ 
Corresponding Theories & String Theory & Doubly Special Relativity \\ 
& & Black Hole Physics  \\
\hline 
\hline 
\end{tabular}
\caption{A comparison between quadratic \cite{guppapers,BHGUP1,BHGUP2,BHGUP3,BHGUP4,BHGUP5,BHGUP6,kmm,kempf}  and linear \cite{Ali:2010yn,Das:2010zf}  GUP approaches in Heisenberg algebra, minimal length uncertainty and maximal moment uncertainty and maximal measured moment \cite{TD2013}. }\label{table1}
\end{center}
\end{table}

\section{Discussion and Conclusions}
\label{sec:concl}

Assuming that the black body behaves as a radiator, Adler calculated the Hawking temperature using Heisenberg uncertainty principle, section \ref{sec:1}. Cavaglia {\it et al.} generalized this in extra dimensions. Modelling the black hole as a sphere of approximately Schwarzschild radius, the energy uncertainty of the emitted photons can be estimated, which is related to the Hawking temperature. The thermodynamic entropy and Bekenstein entropy are calculated. The latter is proportional to the area of the event horizon divided by the Planck area. The rate of photon emission is calculated from the Hawking temperature using Stefan-Boltzmann law, especially when the energy loss is assumed to be dominated by photons. From the mass evaporation, the decay time of the black hole i.e., its lifetime, is given in Eq. (\ref{decayt}). It is apparent that all these quantities would lead to catastrophic evaporation. As discussed in Sec. \ref{ce}, solving that problem using statistical approaches doesn't influence its origin. The present work introduces a recipe of treating the black holes when their temperatures approach the Planckian scale. For instance, we find that the specific heat vanishes only when the black hole entirely evaporates its mass. In other words, the black hole continues the radiation till $m=0$. As the mass approaches zero, the temperature approaches infinity and the radiation rate becomes infinite, as well.

In section \ref{sec:3}, the impacts of GUP on the black hole thermodynamics are discussed. The GUP-approach prevents/protects the black hole from the catastrophic evaporation. The Heisenberg uncertainty principle plays the same role in the hydrogen atom. It prevents it from collapse. We noticed that the GUP assures that considerable remnants will left over, at which the specific heat vanishes. Thus, it is not necessary that the mass entirely evaporates to assure vanishing specific heat. Reaching this stage, the black hole does not exchange heat with its surrounding.

So far, a reliable estimation of black hole lifetime is based on the assumption that it is a radiator. To estimate the black hole lifetime we used another approach, the Salecker-Wigner inequalities, which are assumed to be more severe than the Heisenberg energy-time uncertainty principle. The reason is obvious. The quantum clock is conjectured to show proper time even after the time was being read and the quantum uncertainty in position does not produce a significant inaccuracy in the time measurement. This property is conjuncted to hold over long {\it ''coherent''} time intervals. 

Furthermore, repeating the time measurements does not affect its reliability. Thus the position uncertainty in repeated measurements should be smaller than the minimum wavelength of the reading signals. For an unsqueezed, unentangled and Gaussian signal, the minimum size can be give in minimum mass of the quantum clock. The Wigner first inequality can be applied on black hole, if the size is given by  Schwarzschild radius.

At Planckian scale, the space-time fluctuation become significant. Therefore, the linear spread of quantum clock is to be bound to the Planck distance, for which there is an estimation based on  GUP.  The smallest linear spread of the quantum clock is set to $\alpha_0\, \ell_{p}$. Assuming the mass remains unchanged, the Wigner first inequality is reproduced. When applying GUP-approach, the resulting position uncertainty does not match with Wigner first inequality. The difference depends on the maximum lifetime. Through Wigner second inequality, the latter can be related to the minimum lifetime. 

Assuming that the quantum position uncertainty is limited to the minimum wavelength of measuring signal, the Wigner second inequality can be reproduced. The difference between black hole mass with and without GUP is not negligible. The modified black hole lifetime can be deduced if the spread of quantum clock is limited to a minimum value. The natural one is the Schwarzschild radius. Based on GUP, the resulting lifetime difference depends on black hole mass and $\alpha_0$.

\section*{Acknowledgement}

The author gratefully thanks Ahmed Farag Ali for many discussions and joint work on some topics mentioned in present paper. The author is very grateful to the anonymous referee for drawing his attention to Refs. \cite{swLitr,Zeynali}.


\begin{thebibliography}{99}

\bibitem{wigner57} E. P. Wigner, Rev. Mod. Phys. {\bf 29}, 255 (1957).

\bibitem{wigner58} H. Salecker and E. P. Wigner, Phys. Rev. {\bf 109}, 571 (1958).

\bibitem{disct1} R. J. Adler, I. M. Nemenman, J. M. Overduin and D. I. Santiago, Phys. Lett. B {\bf 477}, 424-428 (2000).

\bibitem{disct2} G. Amelino-Camelia, Phys. Lett. B {\bf 477}, 436-450 (2000).

\bibitem{barrow96} J. D. Barrow,  Phys. Rev. D {\bf 54}, 6563 (1996).

\bibitem{swLitr} Rong-Jia Yang and Shuang Nan Zhang, Phys. Rev. D {\bf 79}, 124005 (2009). 

\bibitem{daxson}  Lance J. Dixon, {\it ''Introduction to conformal field theory and string theory''}, slac-pub-5149, 1989.

\bibitem{guppapers} D. Amati, M. Ciafaloni and G. Veneziano, Phys. Lett. B {\bf 216}, 41 (1989).

\bibitem{BHGUP1} M.~Maggiore,
 Phys.\ Lett.\  B {\bf 304}, 65 (1993),   arXiv:hep-th/9301067.

\bibitem{BHGUP2} M.~Maggiore,
  Phys.\ Rev.\  D {\bf 49}, 5182 (1994), arXiv:hep-th/9305163;
M.~Maggiore,
  Phys.\ Lett.\  B {\bf 319}, 83 (1993),  arXiv:hep-th/9309034.

\bibitem{BHGUP3} L.~J.~Garay, Int.\ J.\ Mod.\ Phys.\  A {\bf 10}, 145 (1995), arXiv:gr-qc/9403008.

\bibitem{BHGUP4} F.~Scardigli,
  Phys.\ Lett.\  B {\bf 452}, 39 (1999),  arXiv:hep-th/9904025.
  
\bibitem{BHGUP5} S.~Hossenfelder, M.~Bleicher, S.~Hofmann, J.~Ruppert, S.~Scherer and H.~Stoecker,
  Phys.\ Lett.\  B {\bf 575}, 85 (2003), arXiv:hep-th/0305262.
  
\bibitem{BHGUP6} C.~Bambi and F.~R.~Urban,
  Class.\ Quant.\ Grav.\  {\bf 25}, 095006 (2008), arXiv:0709.1965 [gr-qc].


\bibitem{kmm} A. Kempf, G. Mangano and R. B. Mann, Phys. Rev. {\bf
D52}, 1108 (1995), arXiv:hep-th/9412167.

\bibitem{kempf} A. Kempf, J.Phys. {\bf A 30}, 2039 (1997), arXiv:hep-th/9604045.

\bibitem{brau} F. Brau, J. Phys. {\bf A 32}, 7691 (1999), arXiv:quant-ph/9905033.

\bibitem{Hossenfelder:2012jw}  S.~Hossenfelder,
   Living Rev. Rel. {\bf 16}, 2 (2013), arXiv:1203.6191 [gr-qc].

\bibitem{advplb} A.~F.~Ali, S.~Das and E.~C.~Vagenas,
  Phys.\ Lett.\  B {\bf 678}, 497 (2009), arXiv:0906.5396 [hep-th].

\bibitem{Ali:2010yn}  A.~F.~Ali, S.~Das and E.~C.~Vagenas,

\bibitem{Das:2010zf}  S.~Das, E.~C.~Vagenas and A.~F.~Ali,
  Phys.\ Lett.\  B {\bf 690}, 407 (2010), arXiv:1005.3368 [hep-th].

\bibitem{sm} J.~Magueijo and L.~Smolin,
  Phys.\ Rev.\ Lett.\  {\bf 88}, 190403 (2002); 
J.~Magueijo and L.~Smolin,
  Phys.\ Rev.\  D {\bf 71}, 026010 (2005), arXiv:hep-th/0401087.

\bibitem{LQG} T. Thiemann, J. Math. Phys. {\bf 39}, 3372-3392 (1998).

\bibitem{Ali:2013ii} A. F. Ali and A. Tawfik, Int. J. Mod. Phys. D {\bf 22}, 1350020 (2013). arXiv:1301.6133 [gr-qc]

\bibitem{Ali:2013ma} A. F. Ali and A. Tawfik, Adv. High Energy Phys. {\bf 2013}, 126528 (2013). arXiv:1301.3508 [gr-qc]

\bibitem{Tawfik:2012he} A. Tawfik, H. Magdy and A.F. Ali, Gen. Rel. Grav. {\bf 45}, 1227-1246 (2013), arXiv:1208.5655 [gr-qc].

\bibitem{Tawfik:2012hz} A. Tawfik, H. Magdy and A.F. Ali, arXiv:1205.5998 [physics.gen-ph]

\bibitem{DahabTaw} E. Abou El Dahab and A. Tawfik, {\it ''On Measurable Maximal Energy and Minimal Time''} in preparation

\bibitem{Elmashad:2012mq} I. Elmashad, A.F. Ali, L.I. Abou-Salem, Jameel-Un Nabi and A. Tawfik, arXiv:1208.4028 [hep-ph].

\bibitem{exp} R. Collela, A. W. Overhauser, and S. A. Werner, Phys. Rev. Lett. 34, 1472 (1975); K. C. Littrell, B. E. Allman, and S. A.
Werner, Phys. Rev. A 56, 1767 (1997); 
  A.~Camacho and A.~Camacho-Galvan,  Rep. Prog. Phys. {\bf 70}, 1-56 (2007).

\bibitem{Pikovski:2011zk}  I.~Pikovski, M.~R.~Vanner, M.~Aspelmeyer, M.~Kim, C.~Brukner, M.~S.~Kim and C.~Brukner,
  Nature Phys.  {\bf 8}, 393 (2012).

\bibitem{dvprl} S. Das and E. C. Vagenas, Phys. Rev. Lett. {\bf 101}, 221301 (2008).

\bibitem{Ali:2011fa}  A.~F.~Ali, S.~Das and E.~C.~Vagenas,
  Phys.\ Rev.\ D {\bf 84}, 044013 (2011).

%
\bibitem{Adler} R.~J.~Adler, P.~Chen and D.~I.~Santiago,
  Gen.\ Rel.\ Grav.\  {\bf 33}, 2101-2108 (2001), arXiv:gr-qc/0106080.

\bibitem{Cavaglia:2003qk} M.~Cavaglia, S.~Das and R.~Maartens,
  Class.\ Quant.\ Grav.\  {\bf 20}, L205-L212 (2003), arXiv:hep-ph/0305223.

\bibitem{Cavaglia1}  
  M.~Cavaglia and S.~Das,
  Class.\ Quant.\ Grav.\  {\bf 21}, 4511-4522 (2004), arXiv:hep-th/0404050.
  
\bibitem{Cavaglia2}   
  A.~J.~M.~Medved and E.~C.~Vagenas,
  Phys.\ Rev.\  D {\bf 70}, 124021 (2004), hep-th/0411022.

\bibitem{Hawking:1974sw}  S.~W.~Hawking,
  Commun.\ Math.\ Phys.\  {\bf 43}, 199-220 (1975).

\bibitem{Bekenstein:1973ur}  J.~D.~Bekenstein,
  Phys.\ Rev.\  D {\bf 7}, 2333 (1973);
  J.~D.~Bekenstein,
  Stud.\ Hist.\ Philos.\ Mod.\ Phys.\  {\bf 32}, 511-524 (2001), arXiv:gr-qc/0009019.

\bibitem{evaporation} R.~Emparan, G.~T.~Horowitz and R.~C.~Myers,
Phys.\ Rev.\ Lett.\  {\bf 85}, 499 (2000), arXiv:hep-th/0003118.

\bibitem{revw} R. Casadio and B. Harms, Entropy {\bf 13}, 502-517 (2011).

\bibitem{TawMBXXXI} A. Tawfik, {\it ''Review in Intensive and Extensive Quantities in HEC''}, invited talk at XXI Max Born Symposium and HIC for FAIR Workshop "Three Days of Critical Behaviour in Hot and Dense QCD", Wroclaw-Poland, 14-16 June 2013

\bibitem{cr1} M. Hotta and M. Yoshimura, Mod. Phys. Lett. A {\bf 9}, 1617-1626 (1994). 

\bibitem{afa:jhep1} A. F. Ali, JHEP {\bf 1209}, 067 (2012). 

\bibitem{hawlt} S. W. Hawking, Nature (London), {\bf 248}, 30 (1974).

\bibitem{TD2013} A. Tawfik and A. Diab, {\it ''Critical Review on Extended and Generalized Uncertainty Principle Scenarios''}, in preparation.

\bibitem{Zeynali} K. Zeynali, F. Darabi, and H. Motavalli, Mod. Phys. Lett. A {\bf 27}, 1250227 (2012), arXiv:1206.5121 [gr-qc].

\bibitem{gupdsr} G. Amelino-Camelia, Int. J. Mod. Phys. D {\bf 11}, 35, (2002); \\
J. Magueijo and L. Smolin, Phys. Rev. Lett. {\bf 88}, 190403, (2002); Phys. Rev. D {\bf 71}, 026010, (2005);\\
J. L. Cortes and J. Gamboa, Phys. Rev. D {\bf 71}, 065015, (2005).


\end{thebibliography}
\end{document}